\documentclass[iop]{emulateapj}

\shorttitle{The Phantom of the Kepler Input Catalog}
\shortauthors{Dalba et al.}
\slugcomment{Accepted for publication in the Astronomical Journal}

\usepackage{subfigure}
\usepackage{url}
\usepackage{hyperref}
\usepackage{epsfig}
\usepackage{graphicx}
\usepackage{sidecap}
\usepackage{hanging}
\usepackage{color}
\usepackage{multirow}
\usepackage{enumerate}
\usepackage{natbib}
\usepackage{amssymb}
\usepackage{amsmath}

\newcommand\pnas{PNAS}
\newcommand{\sci}{Sci}
\newcommand\jatis{JATIS}
\newcommand\msngr{Msngr}

\newcommand{\Kepler}{{\it Kepler} }

\providecommand{\e}[1]{\ensuremath{\times 10^{#1}}}

\begin{document}

\title{\textit{Kepler} Transit Depths Contaminated by a Phantom Star}

\author{Paul A. Dalba\altaffilmark{1,$\dagger$}, Philip S. Muirhead\altaffilmark{1}, Bryce Croll\altaffilmark{2}, and Eliza M.-R. Kempton\altaffilmark{3}}

\affil{\vspace{0pt}\\ $^{1}$Department of Astronomy, Boston University, 725 Commonwealth Ave., Boston, MA, 02215, USA \\
$^{2}$Institute for Astrophysical Research, Boston University, Boston, MA, 02215, USA \\
$^{3}$Department of Physics, Grinnell College, Grinnell, IA 50112, USA}

\altaffiltext{$\dagger$}{\href{mailto:pdalba@bu.edu}{pdalba@bu.edu}}

\begin{abstract}
We present ground-based observations from the Discovery Channel Telescope (DCT) of three transits of \textit{Kepler}-445c---a supposed super-Earth exoplanet with properties resembling GJ 1214b---and demonstrate that the transit depth is $\sim$50\% shallower than the depth previously inferred from \Kepler spacecraft data. The resulting decrease in planetary radius significantly alters the interpretation of the exoplanet's bulk composition. Despite the faintness of the M4 dwarf host star, our ground-based photometry clearly recovers each transit and achieves repeatable 1$\sigma$ precision of $\sim$0.2\% (2 millimags). The transit parameters estimated from the DCT data are discrepant with those inferred from the \Kepler data to at least 17$\sigma$ confidence. This inconsistency is due to a subtle miscalculation of the stellar crowding metric during the \Kepler pre-search data conditioning (PDC). The crowding metric, or CROWDSAP, is contaminated by a non-existent \emph{phantom star} originating in the USNO-B1 catalog and inherited by the \Kepler Input Catalog (KIC). Phantom stars in the KIC are likely rare, but they have the potential to affect statistical studies of \Kepler targets that use the PDC transit depths for a large number of exoplanets where individual follow-up observation of each is not possible. The miscalculation of \textit{Kepler}-445c's transit depth emphasizes the importance of stellar crowding in the \Kepler data, and provides a cautionary tale for the analysis of data from the Transiting Exoplanet Survey Satellite (TESS), which will have even larger pixels than \textit{Kepler}.  
\end{abstract}

\keywords{planets and satellites: fundamental parameters --- planets and satellites: individual (Kepler-445c) --- stars: individual (KIC 9730163, KOI 2704, Kepler-445) --- techniques: photometric}

\section{Introduction}

Among the many discoveries made by the \Kepler Mission \citep{Borucki2010,Koch2010} is the ubiquitous nature of exoplanets with masses and radii in between those of the Earth and Neptune. Despite their absence from the solar system, planets smaller than Neptune, but larger than the solar system's terrestrial planets, are surprisingly common in the Galaxy \citep{Cassan2012,Fressin2013,Petigura2013}. Several subdivisions of this regime have been created (e.g., Super-Earths, mini-Neptunes), but there may exist a continuum of planets in this portion of parameter space, blurring the line that previously separated rocky terrestrial and gaseous giant planets. 

An interesting member of this class of exoplanets is \textit{Kepler}-445c. \textit{Kepler}-445c was first identified as a planet candidate by \citet{Burke2014} and was later confirmed and characterized by \citet{Muirhead2015}. It was reported to be a 2.5$R_{\earth}$ planet orbiting a cool, metal-rich M4 dwarf star $\sim90$ pc away with an orbital period of 4.87 days \citep{Muirhead2015}. The properties of \textit{Kepler}-445c and its host star are similar to the GJ 1214 system, which is one of the most extensively studied exoplanet systems to date. \textit{Kepler}-445 has a \Kepler magnitude of 17.475, making it a relatively faint exoplanet host. This magnitude is slightly above the completeness limit of the \Kepler Input Catalog \citep[KIC,][]{Brown2011} and results in relatively low-precision light curves compared to the majority of \Kepler targets.

Detailed atmospheric investigations have occurred for a handful of exoplanets intermediate in size between the Earth and Neptune \citep[e.g.,][]{Kreidberg2014a,Fraine2014,Knutson2014a,Knutson2014c,Tsiaras2016}, but the vast majority, including \textit{Kepler}-445c, have been classified by their planetary radii inferred from \Kepler transit observations. The light curves returned by \Kepler comprise an invaluable contribution to our understanding of planetary systems, but their analysis is complex and requires a careful assessment of many potential sources of error. The \Kepler Science Operations Center (SOC) pipeline that converts the raw pixel-level data to the simple aperture photometry (SAP) light curves or pre-search data conditioning (PDCSAP) light curves has been thoroughly documented \citep{Jenkins2010,Quintana2010,Bryson2010,Twicken2010,Stumpe2012,Smith2012,Kinemuchi2012}. The PDCSAP light curves undergo extensive artifact mitigation and are well suited for the characterization of transiting exoplanets. 

Because each pixel on the \Kepler detectors covers a 4$\arcsec$x4$\arcsec$ patch of sky and the target apertures typically consist of several pixels, a critical step in the pre-search data conditioning is the correction for crowding, or flux contamination, from sources other than the target star. Indeed, adaptive optics imaging of 3313 \Kepler planet candidate hosts suggests that 14.5$\pm$0.8\% have a nearby star within 0\farcs15 to 4\farcs0 \citep{Ziegler2016}. The susceptibility of the \Kepler data to this source of contamination is well known, and several recent studies have discussed transit dilution by unresolved companion stars \citep[e.g.,][]{Rowe2014,Cartier2015}.

Correcting for crowding requires accurate positions and magnitudes of all the sources in \textit{Kepler's} field-of-view. In most versions of the SOC pipeline, this information comes from the KIC. The crowding correction also requires the expected pixel response functions (PRFs) of the detectors over which the fluxes of the stars will be spread \citep{Bryson2010}. As all of this information from various modules of the pipeline is combined, the chance of an error entering and propagating through the process increases. 

For this reason, many efforts have to been made to remove spurious detections in the \Kepler light curves: \citet{Torres2011} introduced the BLENDER technique to search for blended foreground or background eclipsing binaries diluted by the target, \citet{Morton2012} developed an automated transit candidate validation procedure by calculating false positive probabilities, and \citet{Mullally2016} created a method of identifying false transit signals that were misclassified as small long-period planet candidates.

Here we investigate how a suspicious source in the KIC caused an erroneous crowding correction in the PDCSAP light curves of \textit{Kepler}-445. As a result, the transit depths and planetary radii inferred from the \Kepler data for the exoplanets in this system were overestimated. This investigation was motivated by ground-based observations of several transits of \textit{Kepler}-445c that were obtained with the Discovery Channel Telescope (DCT), which are described in Section \ref{sec:obs}. In Section \ref{sec:analysis}, we describe the analysis of the DCT observations that produced high-precision transit light curves with scatter of 0.17\% (0.0018 mags) in the best case. The repeatable high-precision, time-series photometry from the DCT makes it a valuable tool to be used by the exoplanet community. In Section \ref{sec:analysis}, we also outline the Bayesian parameter estimation technique that yielded transit parameters which were in contention with the previous \Kepler results. In Section \ref{sec:results}, we postulate that an improbable series of events tracing back to the USNO-B1 stellar catalog led to the original mischaracterization of \textit{Kepler}-445c. Finally, in Section \ref{sec:discussion}, we consider the physical radii of all three planets in the \textit{Kepler}-445 system and the potential for additional phantom stars to affect other \Kepler targets. This work offers a cautionary tale for the analysis of data from future exoplanet transit surveys including the Transiting Exoplanet Survey Satellite \citep[TESS,][]{Ricker2015}.

\section{Observations}\label{sec:obs}

We observed three transits of \textit{Kepler}-445c with the Large Monolithic Imager \citep[LMI;][]{Massey2013} on the 4.3-meter Discovery Channel Telescope (DCT) located at Lowell Observatory in Happy Jack, Arizona. The observations are summarized in Table \ref{tab:obs}. Each observation took place under clear conditions and sub-arcsecond seeing. Two transits were observed through the LMI's Sloan i'-band filter on UT 2015 June 19 and UT 2015 July 28 and one transit was observed through the Sloan z'-band filter on UT 2015 July 23.\footnote{LMI transmission curves are available at \url{http://www2.lowell.edu/rsch/LMI/specs.html}.} Each observation sought to obtain at least $\sim$1.2 hours of baseline flux measurements bracketing the $\sim$1.2-hour transit event. Transits were selected so that the entire observation (baseline and transit) occurred while \textit{Kepler}-445 had an airmass below two. 

\begin{deluxetable*}{cccccc}
\tabletypesize{\scriptsize}
\tablecaption{Summary of DCT-LMI Observations of \textit{Kepler}-445 \label{tab:obs}} 
\tablewidth{\textwidth}
\tablehead{ 
\colhead{Date} & \colhead{Filter} & \colhead{Duration} & \colhead{Exposure Time} & \colhead{Airmass} & \colhead{Aperture Radii\tablenotemark{a}} \\
\colhead{[UT]} & & \colhead{[hours]} & \colhead{[seconds]} & & \colhead{[pixels]} } 
\startdata 
2015/06/19 & Sloan i' & 3.87\tablenotemark{b} & 25 & 1.99--1.03 & 12,\;20,\;30  \\  
2015/07/23 & Sloan z' & 3.54\tablenotemark{c} & 14 & 1.02--1.60 & 9.5,\;20,\;30  \\ 
2015/07/28 & Sloan i' & 3.88\tablenotemark{c} & 30 & 1.21--1.02--1.06& 9,\;20,\;30  
\enddata 
\tablenotetext{a}{The values listed in this column correspond to the radii of the photometric aperture, the inner boundary of the sky annulus, and the outer boundary of the sky annulus, respectively.}
\tablenotetext{b}{The final 0.40 hours of the post-transit baseline were excluded from the analysis due to contamination with a partial transit of \textit{Kepler}-445d.}
\tablenotetext{c}{The final 0.17 hours of the post-transit baseline were excluded from the analysis due to a malfunction of the telescope guide camera.}
\end{deluxetable*}

\begin{figure*}
\centering
\includegraphics[scale=0.65]{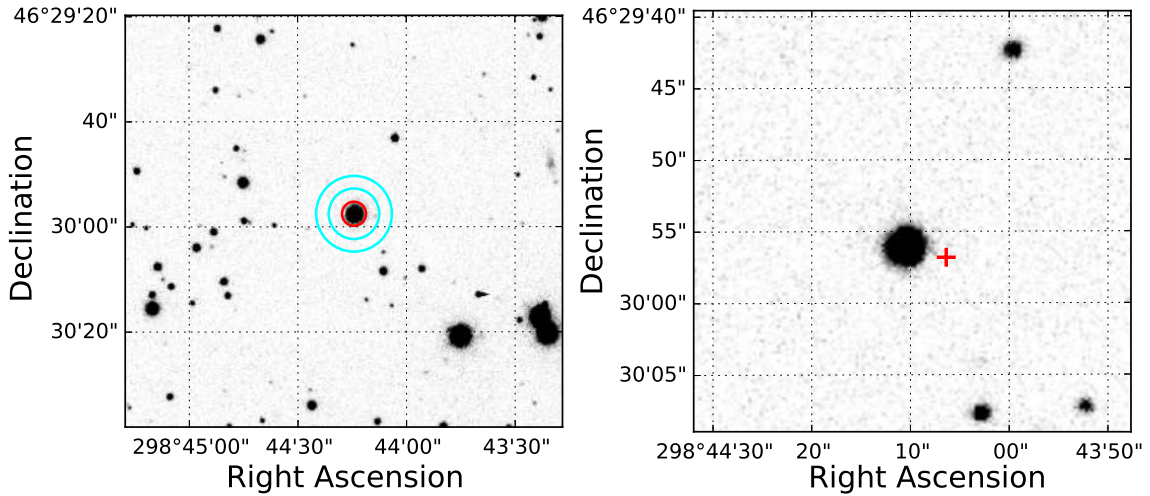}
\caption{\textbf{Left}: Background-subtracted but otherwise raw Sloan z'-band image from the DCT-LMI on UT 2015 July 23 centered on \textit{Kepler}-445. It is only a portion of the full 12$\farcm$3x12$\farcm$3 field-of-view. The minor fringes present in the Sloan z'-band images are difficult to detect on this small scale. The red circle is the 9.5-pixel photometric aperture and the blue circles of radii 20 and 30 pixels define the sky annulus. \textbf{Right}: United Kingdom Infrared Telescope (UKIRT) image centered on \textit{Kepler}-445. Note the difference in scale compared to the DCT image. The red cross denotes the expected position of KIC 9730159 according to the \Kepler Input Catalog. This star was not present in either image.}
\label{fig:fits}
\end{figure*}

We kept \textit{Kepler}-445 in focus during the observations. To ensure that \textit{Kepler}-445 stayed on the same group of pixels, we did not dither in between exposures. With the exposure times listed in Table \ref{tab:obs}, the central pixel of \textit{Kepler}-445's point spread function typically registered $\sim$3\e{4} counts, which was well below the nonlinear response regime of the detector. Figure \ref{fig:fits} shows a portion of a background-subtracted, raw Sloan z'-band image centered on \textit{Kepler}-445.

The LMI has an e2v CCD231 detector, which is a 6144x6160-pixel deep depletion CCD. The readout time for the entire chip with 1x1 binning on a single amplifier is $\sim$73 seconds. Since our exposure times were always less than or equal to 30 seconds, we chose to increase efficiency by reading the chip with all four amplifiers and 2x2 binning. This decreased the overhead time in between exposures to $\sim$8.5 seconds. The effective pixel scale was 0\farcs24 pixel$^{-1}$ with 2x2 binning.

\section{Data Analysis}\label{sec:analysis}

\subsection{Calibration}

We employed a custom data reduction and analysis pipeline that has previously been applied to high-precision, time-series photometry from the DCT-LMI \citep{Dalba2016}. The calibration consisted of a bias and over scan subtraction and a flat-field correction. The minimal dark current present in the observations (estimated to be 0.07 electrons pixel$^{-1}$ hour$^{-1}$) was removed through the background subtraction described below. 

In some cases, we applied an additional correction for fringing. Although the LMI's deep depletion CCD suppressed fringing at long wavelengths, the Sloan z'-band images still displayed low-level fringe patterns with typical variability of less than 1\%. The patterns were large and were difficult to detect in small portions of the full image (Fig. \ref{fig:fits}). We removed the fringing using a fringe frame constructed previously under a separate DCT program (P.I. E. Blanton). The fringe frame was created by median combining 17, 600-second Sloan z'-band images taken on UT 2013 November 02 and UT 2013 November 04. Each image was dithered by at least 15$\arcsec$ to assure that sources fell on different pixels and would not appear in the final fringe frame. Each image also received a bias, over scan, and flat-field correction before being combined to create the fringe frame. 

We scaled the fringe frame to each \textit{Kepler}-445 science frame following the method of \citet{Snodgrass2013} and using 45 ``control pairs'' to account for potential contamination by background sources. The scaled fringe frame was then subtracted from the science frame to remove the fringing.

\subsection{Differential Aperture Photometry}\label{sec:ap_phot}

We conducted differential aperture photometry on \textit{Kepler}-445 using background stars in the 12$\farcm$3x12$\farcm$3 field-of-view of the DCT-LMI. The field was crowded, so we selected calibration stars based on the following criteria. First, the outer edge of sky annulus of the star could not be within 100 pixels of the edge of the frame. Second, the star's sky annulus could not overlap the photometric aperture of any other source in the image. Third, the star's count value could not enter the nonlinear response regime of the detector ($\gtrsim$4\e{4} counts) at any point throughout the observation. Fourth, the calibrator could not be a known variable star. Lastly, the star's photometric aperture could not contain a cosmetic defect (e.g., dead pixels). These criteria yielded hundreds of potential calibration stars. We conducted photometry on each of these stars throughout the observations, although we did not use each star in the calculation of \textit{Kepler}-445's final light curve.

\begin{figure}
\centering
\includegraphics[scale=0.39]{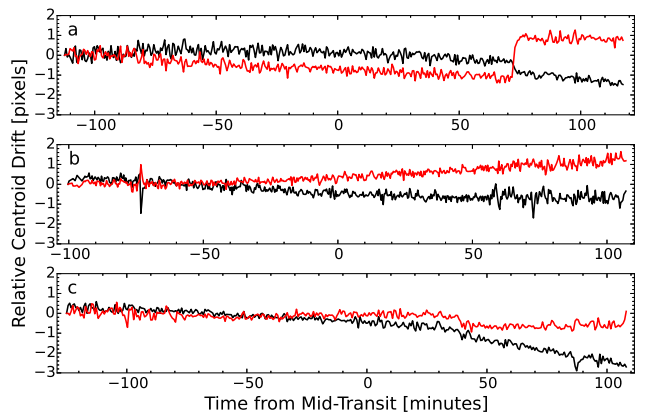}
\caption{Relative drift in centroid positions of \textit{Kepler}-445 during each observation. Panels a, b, and c correspond to observations on UT 2015 June 19, UT 2015 July 23, and UT 2015 July 28, respectively. In each panel, the black curve represents the relative drift of the x-centroid position since the beginning of the observation, and the red curve represents the relative drift of the y-centroid. The centroid jump in panel a resulted from a brief loss of guiding during a reset of the DCT's active optics system.}
\label{fig:centroids}
\end{figure}

For each image, we first determined the centroids of \textit{Kepler}-445 and all calibration stars by fitting a 2D Gaussian profile to the stellar point spread functions. The x- and y-centroid drifts of \textit{Kepler}-445 are shown in Fig. \ref{fig:centroids}. In most cases, the centroid stayed within a pixel or two of its original position. 

We masked each bright source in the image to estimate the median global background value, which was subtracted from the entire image. We summed the flux from \textit{Kepler}-445 and the calibration stars in the photometric apertures and used the sky annuli to account for any residual local background signal, including any dark current. In the sky annuli, bad pixels, hot pixels, cosmic ray strikes, and any other spurious signals at a level of 5$\sigma$ above or below the median count value were masked. We employed this two-stage background subtraction in order to monitor changes in local background across the quadrants of the detector that were read out by different amplifiers. At the time of the \textit{Kepler}-445 observations, the multi-amplifier readout feature of the DCT-LMI had not been widely used. We did not measure local variations in background signal for any of the observations. Since the sky annuli were sufficiently large, the mean local background signals were approximately zero and the second subtraction had no effect on the resultant photometry.

The radii of the photometric apertures and sky annuli are provided in Table \ref{tab:obs}. These apertures yielded the lowest out-of-transit scatter in the final light curves of \textit{Kepler}-445. Other aperture radii in the range [7,14] pixels---incremented by 0.5 pixels---returned less precise photometry.   

Of the many potential calibration stars in the DCT-LMI field-of-view, only a subset was used to create a \emph{master} calibration light curve. The subset was determined through the following procedure. First, each star's light curve was normalized to the median count value of the exposures gathered before or after the transit of \textit{Kepler}-445c. Then, for each of the $i$ stars, we defined a quality factor $Q_i$ as

\begin{equation}
Q_i = \sqrt{\sum_{0}^{N} \;  (F_{445,n} - F_{i,n})^2}
\end{equation}

\noindent where $N$ was the number of exposures taken outside of \textit{Kepler}-445c's transit, $F_{445,n}$ was the median-normalized flux of \textit{Kepler}-445 in the $n$th exposure, and $F_{i,n}$ was the median-normalized flux of the $i$th star in the $n$th exposure. This factor described the out-of-transit deviation of each calibration star's photometry from that of \textit{Kepler}-445. We only continued analysis with the 15 calibration stars having the lowest $Q$-values. These 15 stars were distributed into 32767 unique sets. We calculated a master calibration light curve by taking the mean of all the light curves in each set. We normalized the flux of \textit{Kepler}-445 to each of these 32767 master light curves, and only continued analysis on the one curve that yielded the lowest out-of-transit scatter. The calibration stars used to create final light curves of \textit{Kepler}-445 for each observation are listed in Table \ref{tab:ref}. Each star was similar in brightness to \textit{Kepler}-445, within 1.5 magnitudes in Sloan i'-band.

\begin{deluxetable}{ccc}
\tabletypesize{\scriptsize}
\tablecaption{Final Calibration Stars\label{tab:ref}} 
\tablewidth{\columnwidth}
\tablehead{ 
\colhead{KIC ID\tablenotemark{a}} & \colhead{Sloan i'-band} & \colhead{Observation Date} \\
 & \colhead{Magnitude\tablenotemark{b}} & \colhead{[UT]}}
\startdata 
9789729  & 14.541 & 2015/06/19 \\
9790033  & 14.860 & 2015/06/19 \\
9789938  & 14.913 & 2015/06/19 \\
9789731  & 14.942 & 2015/06/19 \\
9789958  & 14.718 & 2015/07/23 \\
9790304  & 14.557 & 2015/07/23 \\ 
9730187  & 15.493 & 2015/07/23 \\
9730361  & 15.382 & 2015/07/23 \\
9789836  & 15.335 & 2015/07/28 \\
9729892 & 15.733 & 2015/07/28 \\
9729667 & 15.355 & 2015/07/28 \\
9790345 & 15.244 & 2015/07/28 \\
9729595 & 16.540 & 2015/07/28
\enddata 
\tablecomments{The Sloan i'-band magnitude of \textit{Kepler}-445 is 16.024$\pm$0.011 \citep{Muirhead2015}.}
\tablenotetext{a}{\Kepler Input Catalog \citep[KIC,][]{Brown2011}}
\tablenotetext{b}{Magnitudes from \url{http://vizier.u-strasbg.fr/}}
\end{deluxetable}

\begin{figure}
\centering
\includegraphics[scale=0.38]{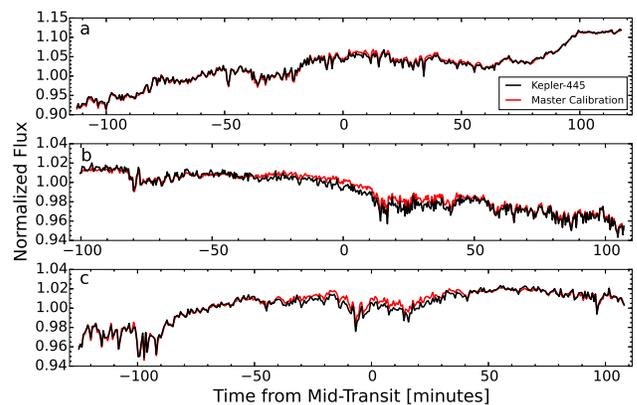}
\caption{Uncalibrated photometry of \textit{Kepler}-445 (black) and the master calibration light curve (red) from each observation. Panels a, b, and c correspond to observations on UT 2015 June 19, UT 2015 July 23, and UT 2015 July 28, respectively. The \textit{Kepler}-445 light curves were normalized to their median out-of-transit values, and the master calibration curves were created following the procedure described in \S\ref{sec:ap_phot}. Note the difference in scale for panel a. Dividing the \textit{Kepler}-445 light curves by the master calibration light curves yielded the photometry in the top panels of Figs. \ref{fig:best_fit_lc_a}, \ref{fig:best_fit_lc_b}, and \ref{fig:best_fit_lc_c}.} 
\label{fig:norm_flux}
\end{figure}

The master calibration light curves and the photometry of \textit{Kepler}-445 prior to calibration are shown in Fig. \ref{fig:norm_flux}. The flux from \textit{Kepler}-445 and the calibration stars increased or decreased as the field rose or set. This variation was greatest for the first observation which sampled the widest range of airmass. The small, rapid variations were due to minor changes in transparency and other noise sources. The transits were visible as deviations between the \textit{Kepler}-445 and calibration light curves. The photometry did not appear to exhibit systematic errors that could influence the values of transit depth derived later.

\begin{figure}
\centering
\includegraphics[scale=0.44]{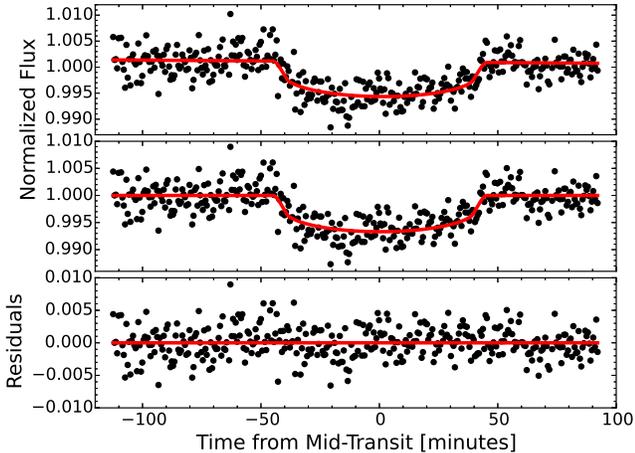}
\caption{Light curves (black points) and best-fit models (red lines) for the transit of \textit{Kepler}-445c on UT 2015 June 19. The top panel shows the data and the model fit including the background signal. In the middle panel, the background signal has been removed from both the model and the data. The bottom panel shows the residuals between the model and the data in the middle panel. A red line is displayed at zero for reference. The 1$\sigma$ precision achieved by these observations is 0.26\%.}
\label{fig:best_fit_lc_a}
\end{figure}

\begin{figure}
\centering
\includegraphics[scale=0.45]{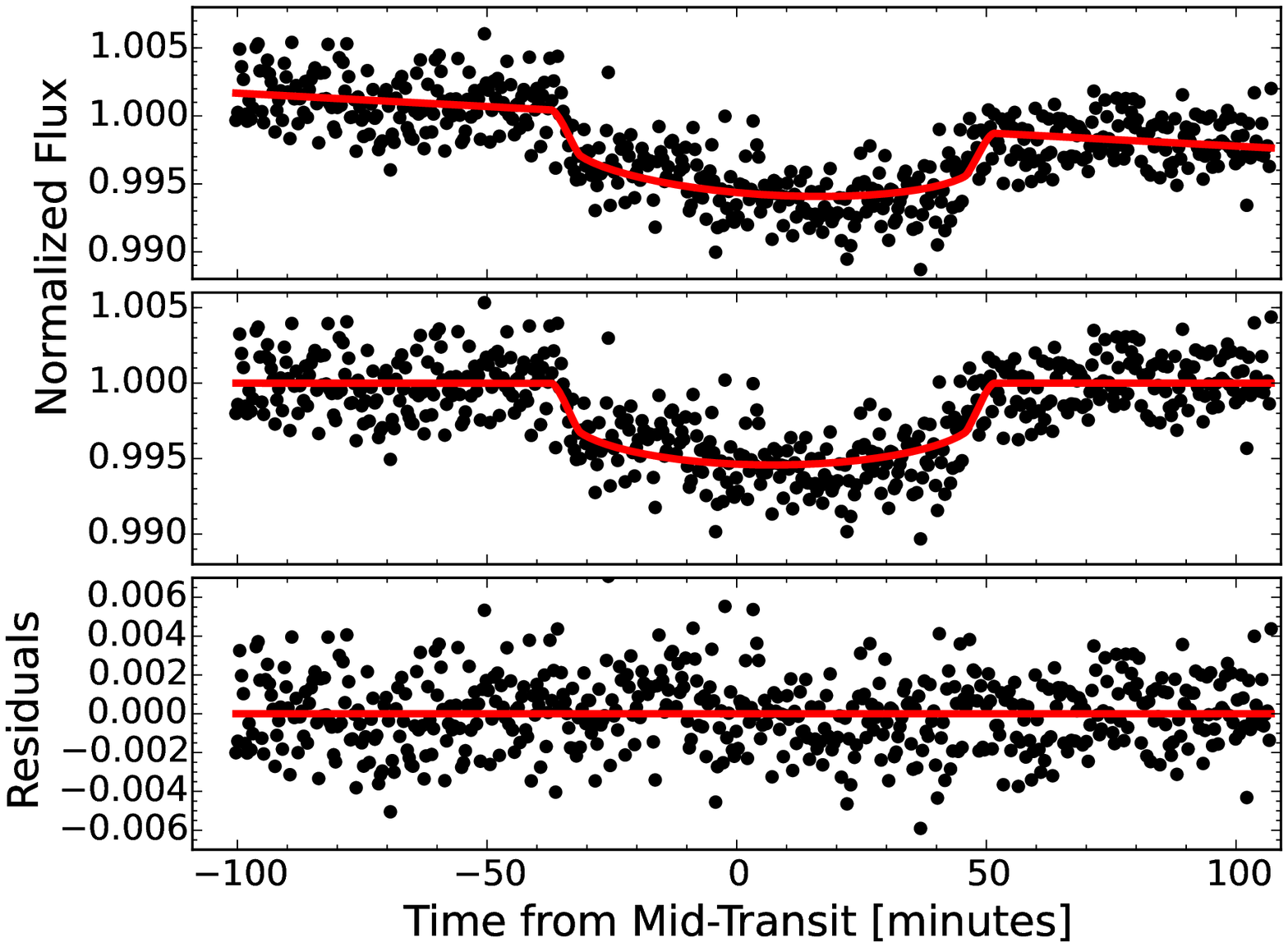}
\caption{Light curves (black points) and best-fit models (red lines) for the transit of \textit{Kepler}-445c on UT 2015 July 23. The description is otherwise identical to Fig. \ref{fig:best_fit_lc_a}. The 1$\sigma$ precision achieved by these observations is 0.23\%.}
\label{fig:best_fit_lc_b}
\end{figure}

\begin{figure}
\centering
\includegraphics[scale=0.45]{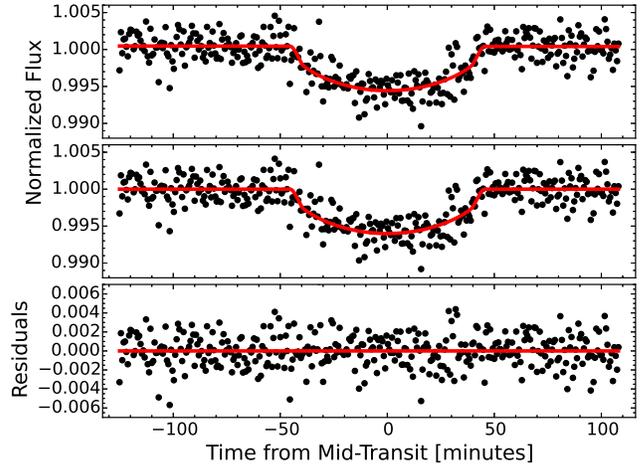}
\caption{Light curves (black points) and best-fit models (red lines) for the transit of \textit{Kepler}-445c on UT 2015 July 28. The description is otherwise identical to Fig. \ref{fig:best_fit_lc_a}. The 1$\sigma$ precision achieved by these observations is 0.17\%.}
\label{fig:best_fit_lc_c}
\end{figure}

The final transit light curves for UT 2015 June 19, UT 2015 July 23, and UT 2015 July 28 are shown in Figs. \ref{fig:best_fit_lc_a}, \ref{fig:best_fit_lc_b}, and \ref{fig:best_fit_lc_c}, respectively. The three observations displayed near-Gaussian scatter with 1$\sigma$ uncertainties of 0.26\%, 0.23\%, and 0.17\%. The repeatable, high-precision photometry demonstrates that the DCT-LMI is a useful tool for transit photometry. 

\begin{figure}
\centering
\includegraphics[scale=0.45]{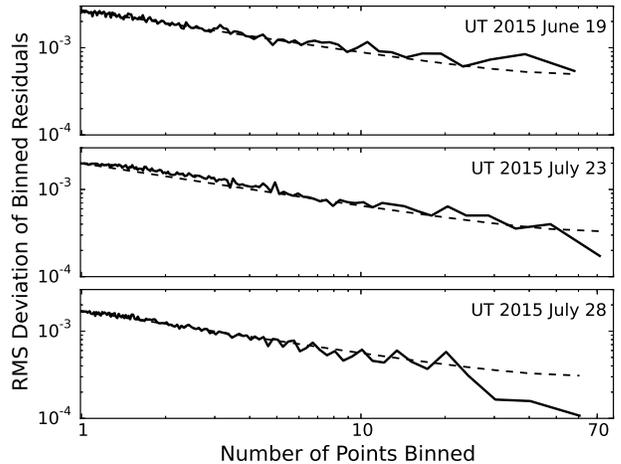}
\caption{Evaluation of time-correlated noise in the \textit{Kepler}-445 photometry using the ``time-averaging'' method of \citet{Pont2006}. For the 3 observations, 10 points corresponded to approximately 5.6, 3.8, and 6.4 minutes, respectively. For each observation, the binned out-of-transit residuals (solid) nearly followed the expectation for Gaussian noise (dashed), suggesting that correlated noise did not greatly influence the photometry.}
\label{fig:rms}
\end{figure}

To date, only a few campaigns have utilized the DCT-LMI for high-precision transit photometry \citep[e.g.,][]{Biddle2014,Dalba2016}, so any time-correlated (red) noise present in this type of observation has not been fully characterized. To evaluate the extent of red noise in the \textit{Kepler}-445 photometry, we employed the ``time-averaging'' method propounded by \citet{Pont2006} and used in numerous transit photometry studies \citep[e.g.,][]{Winn2008}. On the timescale of ingress (5--10 minutes), the binned residuals for each observation stayed very near to the expectation for Gaussian noise (Fig. \ref{fig:rms}). This suggested that our results were not greatly influenced by time-correlated uncertainty. 

We also searched for correlation between the final photometry and the x- and y-centroid positions of \textit{Kepler}-445 on the detector (Fig. \ref{fig:centroids}). Once the background trend was removed from the photometry (\S\ref{sec:param}), the Pearson correlation coefficients ($r$) between the out-of-transit flux and the centroids for each observation were small ($|r|\le0.04$), suggesting that there was no significant linear correlation. 

\subsection{Parameter Estimation}\label{sec:param}

We fit the final light curves of \textit{Kepler}-445 with transit models in order to determine the transit parameters for \textit{Kepler}-445c. The purpose of this parameter estimation was not to fully recharacterize \textit{Kepler}-445c or the \textit{Kepler}-445 system, which we leave to future work (A. Mann et al. 2017, in preparation). Instead, we focused on the magnitude of \textit{Kepler}-445c's transit depth. As we demonstrate in \S\ref{sec:results}, our ground-based observations differed significantly from those obtained with \Kepler data as reported by \citet{Muirhead2015}. 

We employed the analytic transit models of \citet{Mandel2002} in our fits to the DCT-LMI data. The transit parameters used by these models are described in Table \ref{tab:priors}. Following \citet{Kipping2013b}, we included physically realistic quadratic limb darkening in the transit model with the parameters $q_1$ and $q_2$ as substitutes for $u_1$ and $u_2$ via the relations

\begin{equation}
\begin{split}
q_1 & = (u_1 + u_2)^2 \\
q_2 & = \frac{u_1}{2(u_1 + u_2)} \;.
\end{split}
\end{equation}

We also fit for the slight background trend that existed in each observation (see the top panels of Figs. \ref{fig:best_fit_lc_a}, \ref{fig:best_fit_lc_b}, and \ref{fig:best_fit_lc_c}). Although it was possible that this minor effect was astrophysical in nature (i.e., stellar variability), it was more likely a result of the changing airmass during the observations (Table \ref{tab:obs}) and other systematic errors. Regardless of its origin, we included in the fit a background signal ($f_b$) of the form

\begin{equation} \label{eq:bg}
f_b = b_0  + b_1 \, t
\end{equation}

\noindent where $t$ was the time elapsed since the beginning of the observation and $b_0$ and $b_1$ were free parameters. 

Following \citet{Muirhead2015}, we held the orbital eccentricity ($e$) at zero during the model fitting under the assumption that the timescale for \textit{Kepler}-445c to circularize was significantly less than the age of the \textit{Kepler}-445 system. We also adopted the orbital period of \textit{Kepler}-445c from \citet{Muirhead2015} in our fitting procedure since this was well constrained by the \Kepler light curves and not influenced by the post-processing aperture contamination we describe later.   

We used the Markov Chain Monte Carlo (MCMC) ensemble sampler \texttt{emcee} \citep{ForemanMackey2013} to estimate the other transit parameters within a Bayesian framework and applied the uniform priors listed in Table \ref{tab:priors}. Since time-correlated noise was insignificant in these observations (\S\ref{sec:ap_phot}), the uncertainties on each data point were defined to be the standard deviation of the out-of-transit flux. Furthermore, in evaluating the posterior distributions, the likelihood functions were assumed to be Gaussian. Each MCMC routine sampled the parameter space 2\e{6} times, and the chains converged on the best-fit value of each parameter within the first 30\% of the total number of steps. 

The best-fit parameters for each observation fitted individually are presented in Table \ref{tab:ind_fits}. The scaled semi-major axis values ($a/R_{\star}$), which should not vary as a function of wavelength or time, agreed to within 1$\sigma$ across the observations. The limb darkening parameters were not well constrained but marginalizing over these uncertainties still yielded errors less than four percent on the best-fit planet-star radius ratios ($R_p/R_{\star}$). 

The values of $R_p/R_{\star}$ for the two Sloan i' band observations (UT 2015 June 19 and UT 2015 July 28) were inconsistent to $\sim$2.1$\sigma$. The limb darkening parameters found for each of these Sloan i' band observations also displayed notable differences. This may have resulted from the reduction in post-transit baseline observation suffered by the first Sloan i'-band observation. This loss of baseline increased the uncertainty in the estimation of the background trend, potentially allowing the background to artificially influence the transit parameters. The slight difference in observing conditions between the two Sloan i'-band nights may have also exacerbated the issue. The second Sloan i'-band transit, having a full out-of-transit baseline and the highest precision of any observation, was likely the more reliable measurement, and the differences in transit parameters mentioned here were likely not astrophysical in nature.

We extracted the best-fit transit parameters a second time by fitting all three observations simultaneously. The model used in this case had two global parameters ($a/R_{\star}$ and orbital inclination $i$), three filter-dependent parameters ($R_p/R_{\star}$, $u_1$, and $u_2$), and three other local parameters (mid-transit time $t_0$, $b_0$, and $b_1$). The best-fit values of all jointly-fit parameters are listed in Table \ref{tab:joint_fit}. As expected, the new value of $a/R_{\star}$ was consistent with each individual observation. In this joint fit, the values of $R_p/R_{\star}$ between Sloan i' and Sloan z' bands were consistent to within 1$\sigma$. As expected, the Sloan i'-band parameters favored the values from the UT 2015 July 28 observation, which were likely more reliable than those from UT 2015 June 19. 

\begin{deluxetable*}{lcl}
\tabletypesize{\scriptsize}
\tablecaption{Prior Probability Density Functions Employed in all MCMC Fits\label{tab:priors}}
\tablewidth{\textwidth}
\tablehead{
\multicolumn{1}{l}{Parameter} & \colhead{Value or Prior} & \multicolumn{1}{l}{Description}}
\startdata
$e$ & 0 & Orbital eccentricity. Fixed, from \citet{Muirhead2015}  \\
$P$ [days] & 4.871229 $\pm$ 0.000011 & Orbital period. Fixed, from \citet{Muirhead2015}   \\
$a/R_{\star}$  & $\mathcal{U}$[20.41,34.01] & Semi-major axis scaled to stellar radius   \\
$i$ [degrees] & $\mathcal{U}$[88.9,90] & Orbital inclination \\
$t_{0,1}$ - 2457192 [JD] & $\mathcal{U}$[0.75,0.79] & Time of mid-transit on UT 2015 June 19 \\
$t_{0,2}$ - 2457226 [JD] & $\mathcal{U}$[0.85,0.90] & Time of mid-transit on UT 2015 July 23 \\
$t_{0,3}$ - 2457231 [JD] & $\mathcal{U}$[0.73,0.79] & Time of mid-transit on UT 2015 July 28 \\
$q_1$ & $\mathcal{U}$[0,1] & Limb darkening parameter 1 \citep{Kipping2013b} \\
$q_2$ & $\mathcal{U}$[0,1] & Limb darkening parameter 2 \citep{Kipping2013b}\\ 
$R_p/R_{\star}$ & $\mathcal{U}$[0.001,0.01] & Planet-star radius ratio \\
$b_0$ & $\mathcal{U}$[-0.01,0.01] & 0$^{\rm th}$ order background parameter (Eq. \ref{eq:bg}) \\
$b_1$ [days$^{-1}$] & $\mathcal{U}$[-0.05,0.05] & 1$^{\rm st}$ order background parameter (Eq. \ref{eq:bg})
\enddata
\tablecomments{$\mathcal{U}$[$a$,$b$] signifies a uniform prior probability distribution in the range [$a$,$b$].}
\end{deluxetable*}

\begin{deluxetable*}{lccc}
\tabletypesize{\scriptsize}
\tablecaption{Best-Fit Transit Parameters for Individual Observations \label{tab:ind_fits}}
\tablewidth{\textwidth}
\tablehead{
\multicolumn{1}{l}{Parameter} & \colhead{UT 2015 June 19} & \colhead{UT 2015 July 23} & \colhead{UT 2015 July 28}}
\startdata
$a/R_{\star}$        & 26.0$^{+1.0}_{-1.2}$             & 26.2$^{+1.1}_{-1.3}$              & 25.7$^{+1.1}_{-1.4}$  \\
$i$ [degrees]        & 89.50$^{+0.34}_{-0.39}$          & 89.49$^{+0.35}_{-0.39}$           & 89.51$^{+0.34}_{-0.39}$  \\ 
$t_0$ - 2457100 [JD] & 92.76836$^{+0.00063}_{-0.00064}$ & 126.88548$^{+0.00063}_{-0.00055}$ & 131.75565$^{+0.00054}_{-0.00052}$  \\
$q_1$                & 0.55$^{+0.29}_{-0.27}$           & 0.43$^{+0.33}_{-0.23}$            & 0.66$^{+0.23}_{-0.23}$ \\ 
$q_2$                & 0.46$^{+0.31}_{-0.26}$           & 0.51$^{+0.30}_{-0.28}$            & 0.74$^{+0.17}_{-0.20}$ \\
$u_1$                & 0.64$^{+0.34}_{-0.35}$           & 0.63$^{+0.31}_{-0.32}$            & 1.13$^{+0.26}_{-0.26}$ \\
$u_2$                & 0.06$^{+0.42}_{-0.37}$           & -0.02$^{+0.41}_{-0.31}$           & -0.37$^{+0.31}_{-0.22}$ \\
$R_p/R_{\star}$      & 0.0727$^{+0.0026}_{-0.0027}$     & 0.0662$^{+0.0021}_{-0.0023}$      & 0.0651$^{+0.0022}_{-0.0025}$  \\
$b_0$                & 0.00139$^{+0.00031}_{-0.00030}$  & 0.00168$^{+0.00021}_{-0.00021}$   & 0.00047$^{+0.00016}_{-0.00017}$  \\
$b_1$ [days$^{-1}$]  & -0.0045$^{+0.0033}_{-0.0036}$    & -0.0280$^{+0.0024}_{-0.0024}$     & -0.0004$^{+0.0017}_{-0.0016}$ 
\enddata
\tablecomments{The upper and lower uncertainties of each parameter were calculated using the 16th, 50th, and 84th percentiles of the posterior distributions. See Table \ref{tab:priors} for descriptions of the parameters and priors.}
\end{deluxetable*}

\begin{deluxetable}{lc}
\tabletypesize{\scriptsize}
\tablecaption{Best-Fit Transit Parameters for Joint Fit Between All Observations \label{tab:joint_fit}}
\tablewidth{\columnwidth}
\tablehead{
\multicolumn{1}{l}{Parameter} & \colhead{Value} }
\startdata
\textbf{Global} &  \\
\hline \vspace{-0.2cm} \\
$a/R_{\star}$   & 26.2$^{+0.8}_{-1.2}$    \\
$i$ [degrees]   & 89.54$^{+0.32}_{-0.38}$  \vspace{0.2cm}\\
\textbf{Sloan i'} & \\
\hline  \vspace{-0.2cm}\\
$R_p/R_{\star}$ & 0.0675$^{+0.0018}_{-0.0018}$  \\
$q_1$ & 0.55$^{+0.25}_{-0.17}$  \\
$q_2$ & 0.68$^{+0.20}_{-0.21}$ \\
$u_1$ & 0.99$^{+0.20}_{-0.23}$ \\
$u_2$ &-0.27$^{+0.32}_{-0.23}$  \vspace{0.2cm}\\
\textbf{Sloan z'} & \\
\hline  \vspace{-0.2cm}\\
$R_p/R_{\star}$ & 0.0657$^{+0.0019}_{-0.0020}$  \\
$q_1$ & 0.50$^{+0.29}_{-0.22}$ \\
$q_2$ & 0.48$^{+0.30}_{-0.25}$ \\
$u_1$ & 0.65$^{+0.30}_{-0.32}$ \\
$u_2$ & 0.03$^{+0.41}_{-0.35}$ \vspace{0.2cm}\\
\textbf{UT 2015-06-19} & \\
\hline  \vspace{-0.2cm}\\
$t_0$ - 2457100 [JD] & 92.76818$^{+0.00065}_{-0.00065}$\\
$b_0$ & 0.00122$^{+0.00030}_{-0.00030}$ \\
$b_1$ [days$^{-1}$] & -0.0059$^{+0.0035}_{-0.0036}$ \vspace{0.2cm}\\
\textbf{UT 2015-07-23} & \\
\hline  \vspace{-0.2cm}\\
$t_0$ - 2457100 [JD] & 126.88555$^{+0.00057}_{-0.00051}$\\
$b_0$ & 0.00168$^{+0.00021}_{-0.00021}$\\
$b_1$ [days$^{-1}$] & -0.0280$^{+0.0024}_{-0.0024}$ \vspace{0.2cm}\\
\textbf{UT 2015-07-28} & \\
\hline  \vspace{-0.2cm}\\
$t_0$ - 2457100 [JD] & 131.75562$^{+0.00049}_{-0.00048}$ \\
$b_0$ & 0.00055$^{+0.00016}_{-0.00017}$ \\
$b_1$ [days$^{-1}$] & -0.0004$^{+0.0018}_{-0.0016}$ \vspace{0.1cm}
\enddata 
\tablecomments{The upper and lower uncertainties of each fitted parameter were calculated using the 16th, 50th, and 84th percentiles of the posterior distributions. See Table \ref{tab:priors} for descriptions of the parameters and priors.}
\end{deluxetable}

\section{Results}\label{sec:results}

\subsection{The Revised Transit Depth of \textit{Kepler}-445c}

We estimated the transit depths of \textit{Kepler}-445c in the Sloan i' and z' bands using the best-fit parameters in Table \ref{tab:joint_fit}. For each filter, we created 1000 theoretical light curves by drawing samples from the posterior distributions of the transit parameters. From these light curves, we determined the posterior distribution of transit depths. In the Sloan i' and z' bands, these depths were 0.618$\pm$0.022\% and 0.545$\pm$0.028\%, respectively. These values were consistent with each other to 2.1$\sigma$ confidence.

It was unlikely, although not impossible, that the difference between the transit depths was a result of \textit{Kepler}-445c's atmospheric properties. Most exoplanets similar to \textit{Kepler}-445c \citep[e.g., GJ 1214b, HD 97658b, GJ 436b,][]{Kreidberg2014a,Knutson2014a,Knutson2014c} have been found to host atmospheres with high-altitude clouds, hazes, or high mean molecular weights that produce flat transmission spectra. If this was true of \textit{Kepler}-445c's atmosphere as well, one would expect the Sloan i'- and z'-band transit depths to be consistent. The difference, then, could stem from a slight underestimation of the uncertainties on the transit parameters such that the posterior distribution of transit depths was artificially narrow. Regardless of the origin of the inconsistency, the transit depths estimated from the three DCT-LMI observations described here were likely not precise enough to distinguish between various atmospheric compositions for \textit{Kepler}-445c. 

Continuing under the assumption that the variation in transit depths between filters was not astrophysical in nature, we calculated a combined value of transit depth by taking the mean of the two values listed above: 0.582$\pm$0.026\%. The uncertainty was the standard error of the mean defined as $S/\sqrt{n}$ where $S$ was the sample standard deviation and $n$ was the number of samples, which in this case was two.

\subsection{The Previous Transit Depth of \textit{Kepler}-445c}

The \textit{Kepler}-445 system was first characterized with long-cadence \Kepler PDCSAP data by \citet{Muirhead2015}. The $a/R_{\star}$ and $R_p/R_{\star}$ values for \textit{Kepler}-445c reported by that work are 30.21 $\pm$ 0.38 and 0.1075 $\pm$ 0.0014, which are also the values currently listed on the NASA Exoplanet Archive.\footnote{ \url{http://exoplanetarchive.ipac.caltech.edu/}} These values disagree with the $a/R_{\star}$ and $R_p/R_{\star}$ estimated from the DCT observations to 3.2$\sigma$ and greater than 17$\sigma$, respectively. 

Additionally, visual inspection of the phase-folded light curves of \citet[][Figure 5]{Muirhead2015} suggests that the transit depth is at least one percent. \citet{Muirhead2015} does not explicitly report a transit depth, although \url{exoplanets.org} currently cites that work for a depth of 1.16$\pm0.28$\%, more than twice the value reported here.\footnote{Database accessed on 2016 November 21.} Differences in limb darkening can alter transit depths obtained through different filters \citep[e.g.,][]{Knutson2007}, however, this discrepancy seems too large to be attributed entirely to limb darkening. 

Since the \textit{Kepler}-445 system was first characterized, new observations of \textit{Kepler}-445 may result in updated stellar parameters (A. Mann et al. 2017, in preparation). These new parameters may alter the physical characteristics of \textit{Kepler}-445c, but would not explain the differences in transit depth between the \Kepler and DCT-LMI transit curves.

\subsection{The Crowding Metric in the \Kepler Pipeline}\label{sec:crowd}

The discrepancy between the transit depths of \textit{Kepler}-445c is a result of the crowding metric applied by the pre-search data conditioning (PDC) module of the \Kepler pipeline \citep{Stumpe2012,Smith2012}. This metric, defined as ``CROWDSAP'' in the header of the \Kepler light curve files, is the ratio of the target flux to the total flux in the photometric aperture. It is a quarterly-averaged, scalar value between zero and unity. Only the PDCSAP version of the flux reported by \Kepler is adjusted by this crowding factor. 

\begin{figure}
\centering
\includegraphics[scale=0.47]{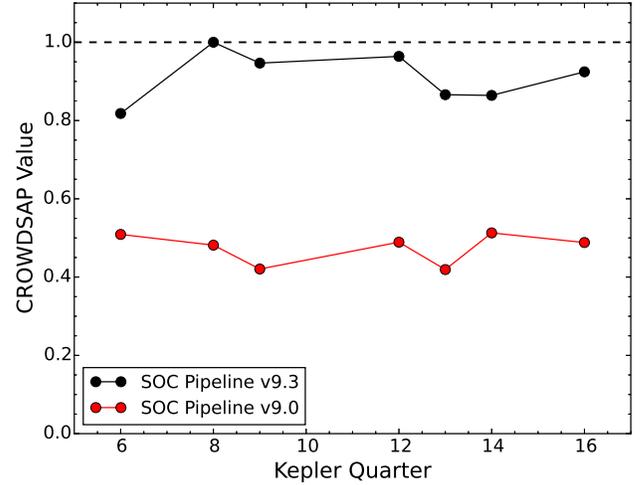}
\caption{The crowding metric for \textit{Kepler}-445 as calculated by two versions of the Science Operations Center (SOC) pipeline. In each case, we ignored the CROWDSAP value from Quarter 17, which was cut short due to the failure of \textit{Kepler's} second reaction wheel. The mean values of CROWDSAP from versions 9.0 and 9.3 were 0.47$\pm$0.04 and 0.91$\pm$0.06, respectively.}
\label{fig:dilution}
\end{figure}

The value of CROWDSAP for a given star may vary from quarter to quarter as it falls on different regions of the detector and is processed with different pixel response functions (PRFs) and optimal photometric apertures. The value of CROWDSAP reported in the \Kepler data also depends on the version of the Science Operations Center (SOC) pipeline that processed the data. In Fig. \ref{fig:dilution}, we show the CROWDSAP values for \textit{Kepler}-445 data processed with versions 9.0 and 9.3 of the SOC pipeline. Between version 9.0, which processed the data used by \citet{Muirhead2015}, and version 9.3, which was released in December 2015,\footnote{See the \Kepler Data Release Notes 25 available on the  Barbara A. Mikulski Archive for Space Telescopes (MAST).} the mean value of CROWDSAP changed from 0.47$\pm$0.04 to 0.91$\pm$0.06, a 6.1$\sigma$ increase. 

CROWDSAP values can be directly affected by changes to the optimal photometric apertures, which occasionally change when the \Kepler pipeline is updated. After an update, the pipeline may determine that a different photometric aperture increases the combined differential photometric precision \citep[CDPP,][]{Christiansen2012} in the resultant light curve of a particular star. If this new aperture now contains flux from nearby sources, then the CROWDSAP parameter should change accordingly.

\begin{figure*}
\centering
\includegraphics[scale=0.6]{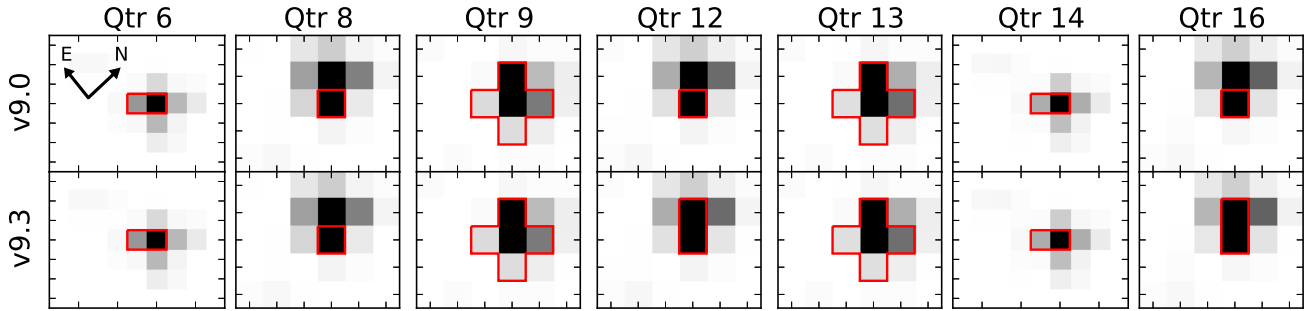}
\caption{Optimal photometric apertures employed by versions 9.0 and 9.3 of the SOC pipeline for \textit{Kepler}-445. The apertures (red polygons) are displayed over target pixel flux data from an arbitrary image taken during that quarter. All frames have the same orientation, and each \Kepler pixel is $\sim4\arcsec$ on a side. The target pixel data were acquired from \Kepler Data Release 25 on MAST, and the same data were displayed for both pipeline versions. The colors were inverted, so pixels with higher count values appeared darker. With the exception of Quarters 12 and 16, the optimal photometric apertures applied in both versions of the SOC pipeline did not change and cannot account for the increased crowding metrics.}
\label{fig:apertures}
\end{figure*}

\begin{figure}
\centering
\includegraphics[scale=0.45]{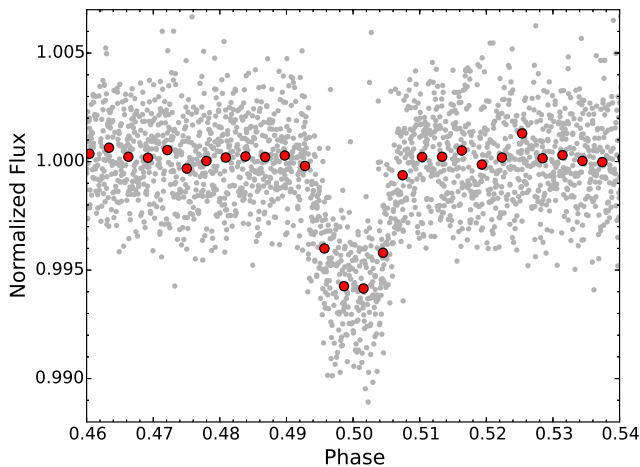}
\caption{Phase-folded, long-cadence \Kepler PDCSAP flux from multiple quarters showing the transit of \text{Kepler}-445c. The data were downloaded from Data Release 25 on MAST---which were processed with version 9.3 of the SOC pipeline---and flattened using the the \texttt{kepflatten} task in PyKE. Mid-transit was set at a phase of 0.5 using the ephemeris of \citet{Muirhead2015}. The gray points are individual exposures and the red points have been binned by a factor of 80. The transit depth is $\sim$0.6\%, clearly less than the $\sim$1\% depth inferred using previous versions of the pipeline. The similarity between this transit curve and the DCT-LMI transit curves supports our theory that the crowding metric caused the discrepancy in transit depth for \textit{Kepler}-445c.}
\label{fig:kep_transit}
\end{figure}

As demonstrated by Fig. \ref{fig:apertures}, the optimal photometric apertures for \textit{Kepler}-445 were largely unchanged between versions 9.0 and 9.3 of the pipeline. With the exception of minor changes in Quarters 12 and 16, the apertures covered the same groups of pixels and yet the CROWDSAP values increased dramatically. From an increase in the average value of the crowding metric, we would expect a corresponding decrease in transit depth. Inspection of the \textit{Kepler}-445c transit curves produced by version 9.3 of the SOC pipeline confirmed this expectation. In Fig. \ref{fig:kep_transit}, we show the phase-folded PDCSAP photometry of \textit{Kepler}-445 from all quarters of observation. The photometry was flattened using the \texttt{kepflatten} task in PyKE\footnote{\url{https://keplerscience.arc.nasa.gov/software.html}} \citep{Still2012} and phase-folded according to the ephemeris in \citet{Muirhead2015} for \textit{Kepler}-445c. Visual inspection of the phase-folded photometry suggested a transit depth of $\sim$0.6\%, which was consistent with the DCT-LMI light curves.

The ratio between the DCT-LMI transit depth of \textit{Kepler}-445c (0.582$\pm$0.026\%) and the previously published transit depth\footnote{See \url{exoplanets.org}, which cites \citet{Muirhead2015}.} (1.16$\pm$0.28\%) equals 0.50$\pm$0.12. This value agrees with the ratio of mean CROWDSAP values between versions 9.0 and 9.3 of the SOC pipeline, 0.52$\pm$0.06, to well within 1$\sigma$. Therefore, the change in the crowding metric offers a highly consistent explanation for the discrepant \textit{Kepler}-445c transit depths discovered by this work. 

The 6.1$\sigma$ increase in mean CROWDSAP value for \textit{Kepler}-445 between the two aforementioned versions of the SOC pipeline results from a new method of constructing the stellar ``scene,'' which contains all stars besides the target and any zodiacal light \citep{Bryson2010}. Before version 9.3 of the SOC pipeline, the scene was determined with the \emph{predicted} PRF model and the stellar data in the KIC. In version 9.3, however, stellar scenes are \emph{reconstructed} through a photometric analysis of the actual pixel level data \citep{Twicken2016}. This new method maximizes the CDPP in the photometric aperture and removes an inherent susceptibility to stellar position and magnitude errors that may be present in the KIC. The optimal photometric apertures may change in response to the new stellar scenes and influence the crowding metrics. However, in the case of \textit{Kepler}-445, the CROWDSAP variations were not caused by changes in photometric apertures.

\subsection{A Phantom Star in the Scene of \textit{Kepler}-445}

The SOC pipeline update and subsequent change in CROWDSAP values identified an unusual circumstance surrounding the characterization of exoplanet \textit{Kepler}-445c. According to the KIC, another source known as KIC 9730159 existed $2\farcs2$ away from \textit{Kepler}-445. The object class of KIC 9730159 was reported as ``star'' and the condition flag was empty (i.e., KIC 9730159 was not flagged as an artifact, planetary-candidate, or exoplanet). KIC 9730159 had a \Kepler magnitude of 17.667, which was 1.1\% greater than that of \textit{Kepler}-445. Given the plate scale of our DCT-LMI observations (0$\farcs$24 pixel$^{-1}$), we would have expected KIC 9730159 to be present in the images $\sim$9 pixels away from \textit{Kepler}-445, well outside the full width at half maximum of the point spread function even during the worst seeing we experienced. Yet KIC 9730159 was not present in the DCT-LMI images (Fig. \ref{fig:fits}). KIC 9730159 was also not visible in the J-band United Kingdom Infrared Telescope (UKIRT)\footnote{Database release ``ukidssdr10plus,'' \url{http://wsa.roe.ac.uk/}} images of the \Kepler field (Fig. \ref{fig:fits}). \citet{Muirhead2015} reported a non-detection of KIC 9730159, postulating that the KIC may have accidentally registered \textit{Kepler}-445 twice. 

We explored the hypothesis that KIC 9730159 was a \emph{phantom star}---present in the KIC and the data conditioning, but not an actual star---in more detail. KIC 9730159 originated in the USNO-B catalog where its designation was 1364-0336533 and its mean epoch of observation was 1990.2. It had a high proper motion (RA:$-146$ mas, DEC:$250$ mas) and a correspondingly large uncertainty in its J2000 coordinates. \citet{Monet2003} warned that objects with large proper motions may be erroneously assigned in future observations and may lead to catalog entries for blank patches of sky. This is a highly probable explanation for the existence of KIC 9730159. \textit{Kepler}-445 (USNO-B1 designation 1364-0336538) also had a relatively high proper motion, but its mean epoch of observation was 1977.1. KIC 9730159 was likely ``created'' as a result of an observation of \textit{Kepler}-445 \textit{circa} 1990 that was mistakenly assigned to a new star. Furthermore, the USNO-B1 catalog reported three detections of this phantom star, allowing it to pass the threshold for acceptance into the KIC \citep{Brown2011}. 

Together, the USNO-B phantom star inherited by the KIC, the original method used to construct the stellar scene of \textit{Kepler}-445, and the timing of the initial characterization of \textit{Kepler}-445c formed an improbable series of events that led to the overestimation of the transit depth of \textit{Kepler}-445c.

\section{Discussion}\label{sec:discussion}

\subsection{Are All the Planets in the Kepler-445 System Rocky?}

The original radius of \textit{Kepler}-445c, 2.51$\pm$0.36$R_{\earth}$ \citep{Muirhead2015}, placed it in the interesting regime of exoplanets with sizes between the Earth and Neptune that are highly amenable to atmospheric characterization \citep[e.g.,][]{Fraine2014,Kreidberg2014a,Knutson2014a,Knutson2014c}. With ground-based observations of additional transits of this exoplanet, we found that the transit depth and planetary radius were overestimated. Using the value of $R_p/R_{\star}$ obtained from the joint fit to all the DCT Sloan i'-band observations\footnote{We choose the $R_p/R_{\star}$ value from Sloan i' band over Sloan z' band based on the parameter uncertainties.} and employing the \textit{Kepler}-445 stellar radius from \citet{Muirhead2015}, we find the revised radius of \textit{Kepler}-445c to be 1.55$\pm$0.23$R_{\earth}$. 

In addition to \textit{Kepler}-445c, this system contains two smaller exoplanets: \textit{Kepler}-445b and \textit{Kepler}-445d. As reported in \citet{Muirhead2015}, the b and d planets have radii of 1.58$\pm$0.23$R_{\earth}$ and 1.25$\pm$0.19$R_{\earth}$, respectively. Although we did not acquire transit observations of these exoplanets, their \Kepler light curves were subject to the same analysis and crowding contamination as that of \textit{Kepler}-445c. Here, we scale their \Kepler transit depths using the ratio of CROWDSAP values before and after the SOC pipeline update (\S\ref{sec:crowd}) and briefly discuss the potential nature of these small exoplanets.

The previously measured transit depths of \textit{Kepler}-445b and \textit{Kepler}-445d are 0.46$\pm$0.36\% and 0.28$\pm$0.58\%, respectively.\footnote{See \url{exoplanets.org}, which cites \citet{Muirhead2015}.} Multiplying these depths by a factor of 0.52 to correct for the phantom star aperture contamination yields depths of $\sim$0.24\% and $\sim$0.15\%. Estimating the transit depth as $(R_{p}/R_{\star})^2$ and applying the \textit{Kepler}-445 stellar radius from \citet{Muirhead2015} yields planetary radii of $\sim 1.1 R_{\earth}$ and $\sim 0.9 R_{\earth}$. The large uncertainties on the transit depths, however, prevent precise estimation of the planetary radii. 

As an alternative means of estimating the planetary radii, we scaled the $R_p/R_{\star}$ values for \textit{Kepler}-445b and \textit{Kepler}-445d to reflect the change in this parameter for \textit{Kepler}-445c discovered by this work. Again employing the \textit{Kepler}-445 stellar radius from \citet{Muirhead2015}, we estimate planetary radii of 0.98$\pm$0.14$R_{\earth}$ and 0.77$\pm$0.12$R_{\earth}$ for \textit{Kepler}-445b and \textit{Kepler}-445d, respectively. 

Despite their Earth-like size, none of the exoplanets in the \textit{Kepler}-445 system orbit within the habitable zone \citep{Muirhead2015}, meaning that none of them are Earth-analogs. It is interesting that the revised radii of all three planets are at or below the threshold of 1.6$R_{\earth}$ from \citet{Rogers2015}, suggesting that all three have primarily rocky compositions. Perhaps, the potentially rocky compositions of short-period exoplanets in systems such as \textit{Kepler}-42 \citep[e.g.,][]{Muirhead2012}, \textit{Kepler}-446 \citep{Muirhead2015}, TRAPPIST-1 \citep{Gillon2016}, and K2-72 \citep[e.g.,][]{Crossfield2016,Vanderburg2016} in addition to \textit{Kepler}-445b, c, and d imply that the formation of small, rocky planets around M-dwarfs is an efficient process. 

However, as is true in all studies of transiting exoplanets, the validity of these conclusions is predicated upon the accuracy of the characterization of the host star, \textit{Kepler}-445. All radii mentioned here rely on the current stellar model \citep{Muirhead2015}, and may change upon further characterization (A. Mann et al. 2017, in preparation).

\subsection{The Potential for Additional Phantom Stars in the KIC}

Every star known (or thought) to exist in the \Kepler field-of-view was assigned a KIC identifier and a \Kepler magnitude. As a result, sources from many catalogs with various levels of accuracy were incorporated into the KIC. Although each source underwent a thorough vetting procedure \citep{Brown2011}, there were bound to be spurious additions, such as KIC 9730159. Since the original characterization of \textit{Kepler}-445c, the updates to the SOC pipeline have increased the accuracy of the CROWDSAP values for all KIC sources. These updates corrected the transit depth of \textit{Kepler}-445c and greatly reduced the probability of errors in the KIC contaminating other \Kepler PDCSAP light curves.

However, it is possible that other phantom stars exist in the KIC and influence the photometry of other \Kepler targets. The authors contributed a discussion on this topic to the 2016 August 8 version of the \Kepler Data Release Notes 25 on MAST.\footnote{Section A.1.2, \url{https://archive.stsci.edu/kepler/data_release.html}} Conducting a direct search for phantom stars is a challenging task given the number of sources comprising the KIC. Indirect searches (e.g., by flux or transit depth changes between different pipeline versions) are also complicated since large CROWDSAP variations can correctly accompany changes in the optimal photometric apertures. Furthermore, different versions of the pipeline typically operate with different amounts of data. The sensitivity added by additional transits must also be considered when searching for phantom stars indirectly.  

Phantom stars are likely a rare occurrence. However, their presence could cause systematic shifts in stellar crowding between pipeline versions and could alter transit depths and inferred planetary radii. This has the potential to affect investigations that make use of the PDCSAP transit depths for a large number of exoplanets, where individual follow-up observation of each is not possible (e.g., studies of planet occurrence rates, planet populations, or transit signal recovery efficiencies).

The sizes of exoplanets in the perplexing Earth-to-Neptune regime only span a few Earth radii. A factor of two correction to the transit depth, corresponding to a $\sim\sqrt{2}$ change in planetary radius, can therefore crucially alter the interpretation of a planet's interior and atmosphere. In this way, assessing the crowding in the scene of an exoplanet host---which is a necessary step in the data conditioning process---has the potential to obscure the understanding of these exoplanets further. Although the case of \textit{Kepler}-445c was highly improbable, it emphasizes the level of caution that must be exercised when analyzing and interpreting data from exoplanet transit surveys.

\subsection{Implications for TESS}

In the near future, the Transiting Exoplanet Survey Satellite (TESS) is expected to launch and begin surveying the entire sky for nearby exoplanets \citep{Ricker2015}. Whereas the pixels on the \Kepler detectors have a plate scale of $\sim4\arcsec$ pixel$^{-1}$, TESS' pixels will cover even more sky, with $\sim21\arcsec$ pixel$^{-1}$. TESS will therefore be even more susceptible to crowding and stellar catalog errors than \textit{Kepler}. The analysis of TESS observations will of course be informed by the many lessons learned from the \Kepler data set, and a careful treatment of the crowding metric should be no exception.

\acknowledgements

The authors thank the referee, Jessie Christiansen, for a thoughtful review that improved the quality of this work. The authors acknowledge Jon Jenkins for a helpful discussion of the \Kepler Science Operations Center Pipeline; Andrew Mann, Trent Dupuy, and Marshall Johnson for information regarding the \textit{Kepler}-445 system; and Andrew Vanderburg for providing a portion of the data used in this paper. The authors also thank Brian Taylor, Stephen Levine, Alan Marscher, Elizabeth Blanton, and the DCT telescope operators for their assistance with the observations. 

These results made use of the Discovery Channel Telescope at Lowell Observatory. Lowell is a private, non-profit institution dedicated to astrophysical research and public appreciation of astronomy and operates the DCT in partnership with Boston University, the University of Maryland, the University of Toledo, Northern Arizona University and Yale University. LMI construction was supported by a grant AST-1005313 from the National Science Foundation. 

This paper includes data collected by the \Kepler Mission. Funding for the \Kepler Mission is provided by the NASA Science Mission Directorate. The \Kepler data presented in this paper were obtained from the Mikulski Archive for Space Telescopes (MAST). STScI is operated by the Association of Universities for Research in Astronomy, Inc., under NASA contract NAS5-26555. Support for MAST for non-HST data is provided by the NASA Office of Space Science via grant NNX13AC07G and by other grants and contracts.

This research made use of the NASA Exoplanet Archive, which is operated by the California Institute of Technology, under contract with the National Aeronautics and Space Administration under the Exoplanet Exploration Program.

This paper includes an image from the United Kingdom Infrared Telescope (UKIRT) Project which is operated by the Joint Astronomy Centre on behalf of the Science and Technology Facilities Council of the United Kingdom.

This work made use of PyKE, a software package for the reduction and analysis of \Kepler data. This open source software project is developed and distributed by the NASA Kepler Guest Observer Office.

This research made use of analytic transit models based on the formalism from Mandel \& Agol (2002) and adapted for \texttt{Python} by Laura Kreidberg.

{\it Facilities:} \facility{Discovery Channel Telescope, (Large Monolithic Imager); \Kepler}

\end{document}